\documentstyle[12pt,epsfig]{article}  
\begin{document} 
\baselineskip=20pt

\def\la{\mathrel{\mathpalette\fun <}}
\def\ga{\mathrel{\mathpalette\fun >}}
\def\fun#1#2{\lower3.6pt\vbox{\baselineskip0pt\lineskip.9pt
\ialign{$\mathsurround=0pt#1\hfil##\hfil$\crcr#2\crcr\sim\crcr}}} 

\begin{titlepage} 
\begin{center}
{\Large \bf Double parton distributions in the leading logarithm approximation 
of perturbative QCD } \\

\vspace{4mm}
 A.M.~Snigirev  \\
M.V.Lomonosov Moscow State University, D.V.Skobeltsyn Institute of Nuclear 
Physics \\
119992, Vorobievy Gory, Moscow, Russia \\ E-mail:~~snigirev@lav01.sinp.msu.ru 
\end{center}  

\begin{abstract} 
Recent CDF measurements of the inclusive cross section for a double parton 
scattering attach a great importance to any theoretical  calculations 
of two-particle distribution functions. Using a parton interpretation of
the leading logarithm diagrams of perturbative QCD theory, 
 generalized Lipatov-Altarelli-Parisi-Dokshitzer
equations for the two-parton distributions are re-obtained. The solutions
of these equations are not at all the product of two single-parton distributions
what is usually applied to the current analysis as ansatz. 
\end{abstract}

\bigskip

\noindent 
$PACS$: ~~12.38.Mh, 24.85.+p, 25.75.+r \\ 
$Keywords$: many parton distributions, leading logarithm  approximation

\end{titlepage}   
\newpage 


The CDF Collaboration has recently measured a large number of double parton
scatterings ~\cite{cdf}. Thus new and complementary information on the structure
of the proton can be obtained by identifying and analyzing events in which two
parton-parton hard scatterings take place within one $p \bar{p}$ collision.
This process, double parton scattering,  provides information on both the spatial
distribution of partons within the proton, and possible parton-parton
correlations. Both the absolute rate for the double parton process and any
dynamics that correlations may introduce are  therefore of interest. 
The theoretical estimations of the effect under consideration 
have been done in a number of works~\cite{goebel, paver, humpert, odorico, 
sjostrand, trelani, del}.

For instance, the
differential cross section for the four-jet process (due to the simultaneous
interaction of two parton pairs) is given by~\cite{humpert, odorico}  
\begin{equation} 
\label{fourjet}
d \sigma = \sum \limits_{q/g} \frac{ d \sigma_{12} ~d \sigma_{34}}
{\sigma_{eff}}~ D_ p(x_1,x_3)~D_{\bar{p}}(x_2,x_4), 
\end{equation} 
$d \sigma_{ij}$ stands for the two-jet cross section. The factor $\sigma_{eff}$
in the denominator represents the total inelastic cross section which is an
estimate for the size of hadron, $\sigma_{eff} ~\simeq~ \pi r_ p^2$. Based on
the simple "hard sphere" model of proton structure, the expected value for  
$\sigma_{eff}$ is 11 mb and is consistent with CDF measured value of 
$(14.5 \pm 1.7^{+1.7}_{-2.3})$ mb~\cite{cdf}. The
two-parton distributions are supposed to be the product of two single-parton
distributions times a momentum conserving phase space factor
\begin{equation} 
\label{factoriz}
D_ p(x_i,x_j)~ = ~ D_ p(x_i,Q^2)~ D_ p(x_j,Q^2)~(1-x_i-x_j),
\end{equation} 
where  $D_ p(x_i,Q^2)$ are the single quark/gluon momentum distributions at
the scale $Q^2$ (deter\-mi\-ned by a hard process).

The main purpose of this Letter is to analyze the status of the factorization  
ansatz
(\ref{factoriz}) for many parton distributions in the perturbative QCD theory.

Here one should note that the generalized Lipatov-Altarelli-Parisi-Dokshitzer
equations for many parton distribution functions have been derived for the first
time in refs.~\cite{kirschner, snig} within the leading logarithm approximation
of QCD using a method by Lipatov~\cite{lipatov}.
Under certain initial conditions these
equations lead to the solutions, which are identical with the jet calculus rules
proposed for multiparton fragmentation functions by
Konishi-Ukawa-Veneziano~\cite{konishi}. Because of a very old affair it is
necessary to recall some features of that investigation to be clear.  

In ref.~\cite{gribov} the structure functions of $ep$ scattering and
$e^+e^-$ annihilation were calculated in the leading logarithmic approximation
for vector and pseudoscalar theories. Similar calculations in QCD were made
in~\cite{dok}. Lipatov shown~\cite{lipatov} that the results of these
calculations admit a simple interpretation in the framework of the parton model
with a variable cut-off parameter $\Lambda~\sim~ Q^2$ with respect to the
transverse momenta, and derived an equation for the scaling violation
of the parton distribution $ D^j_i(x,\Lambda)$ inside a dressed quark or gluon,
which is in fact equivalent to the one proposed by Altarelli and
Parisi~\cite{altarelli}, within the difference that it was not applied by
Lipatov to QCD.

Calculating ladder graph contributions to the structure functions it is
convenient to use Sudakov's technique~\cite{sudakov}:
\begin{equation}
\label{1.1}
k=\alpha q{'} + \beta p
+k_{\perp},~q{'}=q-\frac{q^2}{s}p,~q{'}^2=0,~pq{'}=\frac{1}{2}s,
\end{equation}
\noindent $q$ is the momentum of the virtual photon, $p$ is the momentum
($p^2=0$) of the considered quark or gluon. In deep inelastic processes this
constituent is one of the partons found at the reference virtualness
$-q^2_0=\mu^2$ in the target proton. In $e^+e^-$ annihilation it is that final
quark or gluon which converts into the detected hadron. $q^{'}$ is a light-like
vector combined from $p$ and $q$. $k_{\perp}$ is the component of $k$ orthogonal
to $q^{'}$ and $p$. In the next step one takes the residues with respect to
$\alpha$ in the Feynman integrals. Now each line $r$ carries only $\beta_r$
and $k_{\perp r}$. And so one may introduce the amplitudes 
$\Psi_i^n(\beta_{r},~k_{\perp r})$ for finding in the dressed constituent 
$i$~~$n$ bare partons of given types and parameters $\beta_{r},~k_{\perp r}$, 
$r=1,~...,~n$. Analogously
$\bar \Psi_i^n (\beta_r,~k_{\perp r})$ are the probability amplitudes for
finding a bare parton of $i$-type in a state of $n$ dressed partons with Sudakov
parameters $\beta_{r},~k_{\perp r}$ in the case of $e^+e^-$ annihilation. These
amplitudes can be calculated as tree diagrams in an old-fashioned perturbation
theory and obey the normalization condition
\begin{equation}
\label{normalization}
1= z_i + \sum\limits_{n=2}^{\infty} \int \prod\limits_{r=1}^n
\frac{d\beta_r}{\beta_r}\theta(\beta_r)d^2k_{\perp r} |\Psi_i^n|^2 \delta^2(\sum
k_{\perp r}) \delta(\sum \beta_r -1),
\end{equation}
\noindent
where $z_i$ is the wave function renormalization constant, to be interpreted as
the probability that the constituent $i$ remains in the bare state.

An important feature of the leading logarithm diagrams is the ordering of the
lines. The absence of interference-type diagrams allows a classical probability
interpretation. The integration region in $k_{\perp}$ essential in the leading
logarithm approximation goes up to $\Lambda =|q^2|,~|k_{\perp}| < \Lambda $. The
same cut-off is used for the ultraviolet-divergent $z_i$. The unrenormalized
amplitudes $\Psi_i^n$, $\bar \Psi_i^n$ are connected with the renormalized 
(cut-off independent)
ones $\varphi_i^n$, $\bar \varphi_i^n$ via
\begin{eqnarray}
\label{1.3}
\Psi_i^n & = &\varphi_i^n \prod\limits_j z_j^{n_j/2}(\Lambda), \nonumber \\
\bar \Psi_i^n & = &\bar \varphi_i^n z_i^{1/2}(\Lambda),
\end{eqnarray}
\noindent
$n_j$ being the number of partons $j$ in the state $n$.

In the normalization condition (\ref{normalization}) the left-hand side does not
depend on $\Lambda$, while on the right-hand side the functions $\Psi_i^n$
and the limits of integration over $k_{\perp}$ depend on $\Lambda$.
Differentiating (\ref{normalization}) with respect to $\Lambda$ one obtains
(see refs.~\cite{lipatov, kirschner, snig} for details)
\begin{equation}
\label{zi}
0 = \Lambda \frac{dz_i}{d\Lambda}z_i^{-1} + \omega_i(g_{\Lambda}^2),
\end{equation}
\noindent
where 
\begin{equation}
\label{pi}
 \omega_i(g_{\Lambda}^2)= \frac{ \partial~ \Pi_i(\Lambda/k^2)}{\partial
 \ln(\Lambda/|k^2|)}\Bigg
 |_{k^2=\Lambda}~=~\frac{g_{\Lambda}^2}{8\pi^2}\bar{\omega_i},
\end{equation}
\noindent
$\Pi_i$ is the one-loop self-energy part of the constituent $i$, ~$g_{\Lambda}$
is the running coupling constant. In eq. (\ref{pi}) one differentiates only the
upper limit of the loop integral.

The parton distribution function can be defined in this framework as 
\begin{eqnarray}
\label{singl}
D_i^j(x,\Lambda) =  z_i \delta_{ij} \delta(x-1) + \sum\limits_{n=2}^{\infty}
\sum\limits_{r(j)} \int \prod\limits_{r=1}^n
\frac{d\beta_r}{\beta_r}\theta(\beta_r)d^2k_{\perp r}\cdot \nonumber \\
\cdot |\Psi_i^n|^2 \delta^2(\sum k_{\perp r}) \delta(\sum \beta_r-1)
\delta(\beta_{r(j)}-x). 
\end{eqnarray}
\noindent
The second sum $ \sum \limits_{r(j)}$ runs over all partons of type $j$ in the 
state $n$. The result of differentiation reads
\begin{equation}
\label{esingl}
\Lambda \frac{dD_i^j(x,\Lambda)}{d\Lambda} = \frac{g_{\Lambda}^2}{8\pi^2}
\sum\limits_{j{'}} \int \limits_x^1
\frac{dx{'}}{x{'}}D_i^{j{'}}(x{'},\Lambda)P_{j{'}\to j}\Bigg(\frac{x}{x{'}}\Bigg),
\end{equation}
\noindent
where
$$\frac{g_{\Lambda}^2}{8\pi^2} \frac{1}{x} P_{j \to j_1}
\Bigg(\frac{x_1}{x}\Bigg) =
\omega_{j \to j_1}(x \to x_1) -
\delta_{jj_1}\delta(x-x_1)\omega_j(g_{\Lambda}^2),$$
\begin{equation}
\label{1.8}
\omega_{j \to j_1} = \sum\limits_{j_1 \leq j{'}} \omega_{j \to j_1j{'}} +
\sum\limits_{j{'} \leq j_1} \omega_{j \to j{'}j_1},~~~~~\omega_j =
\sum\limits_{j_1 \leq j_2} \omega_{j \to j_1j_2}, 
\end{equation}
\noindent
$\omega_{j \to j_1j_2} (x \to x_1) dx_1 d\Lambda /\Lambda$, as usual, is the
probability that a parton of type $j$ with the fraction $x$ of the longitudinal
momentum decays into two partons of types $j_1$ and $j_2$, one of which has the
fractions $x_1$ of the longitudinal momentum and the transverse momentum
$\Lambda$. These probabilities are defined as in eq.(\ref{pi}) but with the
types of partons and the longitudinal momentum fractions in the loop specified.
Also, the eq.(\ref{zi}) was used to derive the evolution equation (\ref{esingl}).

By introducing the natural variable
$$t = \frac{1}{2\pi b} \ln \Bigg[1 + \frac{g^2(\mu^2)}{4\pi}b
\ln\Bigg(\frac{\Lambda}{\mu^2}\Bigg)\Bigg],~~~~~b = \frac{33-2n_f}{12\pi}~~
{\rm {in~ QCD}},$$
\noindent
one obtains
\begin{equation}
\label{e1singl}
 \frac{dD_i^j(x,t)}{dt} = 
\sum\limits_{j{'}} \int \limits_x^1
\frac{dx{'}}{x{'}}D_i^{j{'}}(x{'},t)P_{j{'}\to j}\Bigg(\frac{x}{x{'}}\Bigg).
\end{equation}
\noindent
This is an equation in the form of Altarelli and  Parisi~\cite{altarelli}. It is
interesting that expression (\ref{1.8}) for the kernels $P$ already includes a
regularization at $x \rightarrow x{'}$, which was introduced in
ref.~\cite{altarelli} afterwards.

Defining the two-parton distribution function
\begin{eqnarray}
\label{double}
D_i^{j_1j_2}(x_1,x_2,\Lambda)  = 
z_i\delta_{ij_1}\delta_{j_1j_2}\delta(x_1-x_2)\delta(x_1-1)
-\delta_{j_1j_2}\delta(x_1-x_2)D_i^{j_1}(x_1,\Lambda)+\nonumber\\
+ \sum\limits_{n=2}^{\infty} \sum\limits_{r(j_1)} \sum\limits_{r(j_2)} \int
\prod\limits_{r=1}^n \frac{d\beta_r}{\beta_r} 
\theta (\beta_r)d^2k_{\perp r} |\Psi_i^n|^2 \delta^2(\sum k_{\perp r})
\delta(1-\sum \beta_r)\delta(\beta_{r(j_1)}-x_1) \delta(\beta_{r(j_2)}-x_2), 
\end{eqnarray}
\noindent
one obtains by differentiating with respect to $\Lambda$
\begin{eqnarray}
\label{edouble}
& &\frac{dD_i^{j_1j_2}(x_1,x_2,t)}{dt} =\sum\limits_{j_1{'}}
\int\limits_{x_1}^{1-x_2}\frac{dx_1{'}}{x_1{'}}D_i^{j_1{'}j_2}(x_1{'},x_2,t)
P_{j_1{'}
\to j_1} \Bigg(\frac{x_1}{x_1{'}}\Bigg)+\\
& &+ \sum\limits_{j_2{'}}\int\limits_{x_2}^{1-x_1}
\frac{dx_2{'}}{x_2{'}}D_i^{j_1j_2{'}}(x_1,x_2{'},t)P_{j_2{'} \to j_2}
\Bigg(\frac{x_2}{x_2{'}}\Bigg) 
+ \sum\limits_{j{'}}D_i^{j{'}}(x_1+x_2,t) \frac{1}{x_1+x_2}P_{j{'} \to
j_1j_2}\Bigg(\frac{x_1}{x_1+x_2}\Bigg),\nonumber 
\end{eqnarray}
\noindent 
where the kernel
\begin{equation}
\label{kernel}
\frac{g_{\Lambda}^2}{8\pi^2}\frac{1}{x} P_{j \to
j_1j_2}\Bigg(\frac{x_1}{x}\Bigg) = \omega_{{j \to j_1j_2} \atop {j_1 \leq j_2}}
(x \to x_1) + \omega_{{j \to j_1j_2} \atop {j_1 \geq j_2}}
(x \to x_1)
\end{equation}
\noindent
is defined without $\delta$-function regularization. This is the generalized
Lipatov-Altarelli-Parisi-Dokshitzer equation for two parton distributions
$D_i^{j_1j_2}(x_1,x_2,t)$, representing the probability that in a dressed
constituent $i$ one finds two bare partons  of  types $j_1$ and $j_2$ with the given
momentum fractions $x_1$ and $x_2$ . The
result for the $m$-parton functions can be found in ref.~\cite{kirschner}.

It is readily  verified by  direct substitution that the solution 
of eq.~(\ref{edouble}) can be
written via the convolution of single distributions~\cite{kirschner, snig}
\begin{eqnarray}
\label{solution}
& D_i^{j_1j_2}(x_1,x_2,t) = \\
& \sum\limits_{j{'}j_1{'}j_2{'}} \int\limits_{0}^{t}dt{'}
\int\limits_{x_1}^{1}\frac{dz_1}{z_1}
\int\limits_{x_2}^{1-x_1}\frac{dz_2}{z_2}~
D_i^{j{'}}(z_1+z_2,t{'}) \frac{1}{z_1+z_2}P_{j{'} \to
j_1{'}j_2{'}}\Bigg(\frac{z_1}{z_1+z_2}\Bigg) D_{j_1{'}}^{j_1}(\frac{x_1}{z_1},t-t{'}) 
D_{j_2{'}}^{j_2}(\frac{x_2}{z_2},t-t{'}).\nonumber
\end{eqnarray}
\noindent 
This coincides with the jet calculus rules~\cite{konishi} proposed originally
for the fragmentation functions and is the  generalization of 
 well-known Gribov-Lipatov relation
installed for single functions~\cite{gribov, dok} (the distribution
of bare partons inside a dressed constituent  is identical to the distribution 
of dressed constituents in the fragmentation of 
a bare parton in the leading logarithm 
approximation). The equations for the multiparton fragmentation functions are
obtained by Lipatov's method in a similar way~\cite{snig} and beyond the given
investigation.

The solution (\ref{solution}) shows that the distribution of partons is 
{\it {correlated}} in the leading logarithm approximation:
\begin{eqnarray}
\label{nonfact}
D_i^{j_1j_2}(x_1,x_2,t) \neq D_{i}^{j_1}(x_1,t) 
D_{i}^{j_2}(x_2,t).
\end{eqnarray}
Of course, it is interesting to find out the phenomenological issue of the
equations under consideration. It can be done within the well-known
factorization of soft and hard stages (physics of short and long
distances)~\cite{collins}. As result the equations (\ref{e1singl}) and
(\ref{edouble}) describe the evolution of parton distributions in a hadron with
$t ~(Q^2)$, if one replaces the index $i$ by index $h$ only. However, the initial
conditions for new equations at $t=0 ~(Q^2=\mu^2)$ are unknown a priori and must
be introduced phenomenologically or must be extracted from experiments 
or some models dealing with physics of long distances (at the parton level: 
$D_{i}^{j}(x,t=0)~= ~\delta_{ij} \delta(x-1)$; ~$D_i^{j_1j_2}(x_1,x_2,t=0)~=~0$).
Nevertheless the solution of eq.~(\ref{edouble}) with the given initial
condition may be written as before via the convolution of single
distributions~\cite{snig}
\begin{eqnarray}
\label{solution1}
& D_h^{j_1j_2}(x_1,x_2,t) = 
\sum\limits_{j_1{'}j_2{'}} 
\int\limits_{x_1}^{1}\frac{dz_1}{z_1}
\int\limits_{x_2}^{1-x_1}\frac{dz_2}{z_2}~
D_h^{j_1{'}j_2{'}}(z_1,z_2,0) D_{j_1{'}}^{j_1}(\frac{x_1}{z_1},t) 
D_{j_2{'}}^{j_2}(\frac{x_2}{z_2},t) ~+ \\
& \sum\limits_{j{'}j_1{'}j_2{'}} \int\limits_{0}^{t}dt{'}
\int\limits_{x_1}^{1}\frac{dz_1}{z_1}
\int\limits_{x_2}^{1-x_1}\frac{dz_2}{z_2}~
D_h^{j{'}}(z_1+z_2,t{'}) \frac{1}{z_1+z_2}P_{j{'} \to
j_1{'}j_2{'}}\Bigg(\frac{z_1}{z_1+z_2}\Bigg) D_{j_1{'}}^{j_1}(\frac{x_1}{z_1},t-t{'}) 
D_{j_2{'}}^{j_2}(\frac{x_2}{z_2},t-t{'}).\nonumber
\end{eqnarray}

The reckoning for unsolved confinement problem (physics of long distances) is
unknown two-parton correlation function $ D_h^{j_1{'}j_2{'}}(z_1,z_2,0)$
at some scale $\mu^2$. One can suppose that this function is the product of two
single-parton distributions times a momentum conserving 
factor at this scale $\mu^2$:
\begin{eqnarray}
\label{fact}
D_h^{j_1j_2}(z_1,z_2,0) ~=~ D_{h}^{j_1}(z_1,0) 
D_{h}^{j_2}(z_2,0)\theta(1-z_1-z_2),
\end{eqnarray}
\noindent
then
\begin{eqnarray}
\label{solution2}
& D_h^{j_1j_2}(x_1,x_2,t) = 
\bigg(D_{h}^{j_1}(x_1,t)D_{h}^{j_2}(x_2,t)~+~\\
& \sum\limits_{j_1{'}j_2{'}} 
\int\limits_{x_1}^{1}\frac{dz_1}{z_1}
\int\limits_{x_2}^{1}\frac{dz_2}{z_2}~
D_h^{j_1{'}}(z_1,0)D_h^{j_2{'}}(z_2,0)
 D_{j_1{'}}^{j_1}(\frac{x_1}{z_1},t) 
D_{j_2{'}}^{j_2}(\frac{x_2}{z_2},t)[\theta(1-z_1-z_2)-1]
\bigg)\theta (1-x_1-x_2)~+
\nonumber \\
& \sum\limits_{j{'}j_1{'}j_2{'}} \int\limits_{0}^{t}dt{'}
\int\limits_{x_1}^{1}\frac{dz_1}{z_1}
\int\limits_{x_2}^{1-x_1}\frac{dz_2}{z_2}~
D_h^{j{'}}(z_1+z_2,t{'}) \frac{1}{z_1+z_2}P_{j{'} \to
j_1{'}j_2{'}}\Bigg(\frac{z_1}{z_1+z_2}\Bigg) D_{j_1{'}}^{j_1}(\frac{x_1}{z_1},t-t{'}) 
D_{j_2{'}}^{j_2}(\frac{x_2}{z_2},t-t{'}),\nonumber
\end{eqnarray}
where
\begin{equation}
\label{1solution}
 D_h^j(x,t) = 
\sum\limits_{j{'}} \int \limits_x^1
\frac{dz}{z}~D_h^{j{'}}(z,0)~D_{j{'}}^j(\frac{x}{z},t)
\end{equation}
\noindent
is the solution of eq.~(\ref{e1singl}) with the given initial condition
$D_h^j(x,0)$ for parton distributions inside a hadron expressed via
distributions at the parton level. 

This result~(\ref{solution2}) is, as matter of fact, the answer to the question
set. If the two-parton distributions are factorized at some scale $\mu^2$ then
the evolution violates this factorization {\it{ inevitably}} at any different
scale ($Q^2 \neq \mu^2$), apart from the violation due to 
the kinematic correlations induced by the momentum
conservation (given by $\theta$-functions){\footnote{ Here this is the analog of
a momentum conserving phase space factor in eq.~(\ref{factoriz})}}.

For a practical employment it is interesting to know the degree of this
violation. It can be done numerically using, for instance, the
CTEQ-parametrization~\cite{cteq} for single distributions as an input in
eq.~(\ref{solution2}) and considering the kinematics of some specific process.
Partly this problem was investigated theoretically in refs.~\cite{snig, snig2}
and  for the two-particle correlations of fragmentation
functions in ref.~\cite{puhala}. That technique is based on the Mellin 
transformation of distribution functions like
\begin{eqnarray}
\label{mellin}
M_h^{j}(n,t) ~=~ \int\limits_{0}^{1}dx ~x^n~D_{h}^{j}(x,t). 
\end{eqnarray}
\noindent
After that the integro-differential equations (\ref{e1singl}) and
(\ref{edouble}) become systems of ordinary linear-differential equations of
first order with constant coefficients and can be solved explicitly~\cite{snig,
snig2}. In order to obtain the distributions in $x$-representation an inverse
Mellin transformation must be performed 
\begin{eqnarray}
\label{mellin in}
D_h^{j}(x,t) ~=~ \int\frac {dn}{2\pi i} ~x^{-n}~M_{h}^{j}(n,t), 
\end{eqnarray}
\noindent
where the integration runs along the imaginary axis to the right from all
$n$-singularities. This can be done numerically again. However the asymptotic
behaviour can be estimated. Namely, with the growth of $t~(Q^2)$ the second term in
eq.~(\ref{solution1}) becomes {\it {dominant}} for the finite $x_1$ 
and $x_2$~\cite{snig2}.  
Thus the two-parton distribution functions 
"forget" the initial conditions unknown a
priori and the correlations perturbatively calculated appear.

The CDF Collaboration found no evidence for the kinematic correlation between
the two scatterings in double parton events. This can mean only that the
factorization ansatz (\ref{factoriz}) is the acceptable approximation at the
scale $Q_{CDF} \sim 5$ GeV accessible to CDF measurements ($ E_T^{jet}(min) 
\simeq 5$ GeV~\cite{cdf}). There are no arguments  to assert that this ansatz
(\ref{factoriz}) is acceptable at the larger scales of hard processes accessible
to LHC measurements. The second term in eqs.~(\ref{solution1}) and 
(\ref{solution2}) can give the meaning contribution to the cross section of many
jet production at LHC energy and influence on the background estimations to the
Higgs production.

 To summarize, the analysis shows that within the leading logarithm
 approximation of the perturbative QCD theory and the factorization of physics
 of short and long distances, the two-parton distribution functions being the
 product of  two single distributions at some reference scale become to be
 dynamically correlated at any different scale of 
 a hard process. These correlations are
 perturbatively calculable (\ref{solution2}). Of course, in order to be more
 conclusive one needs to do numerical estimations of this effect that is
 planning to perform in the future.

\bigskip

\noindent
{\it Acknowledgements}

It is pleasure to thank V.A. Ilyin drawing the attention to
the problem of a double parton scattering.
Discussions with E.E.~Boos, I.P. Lokhtin, S.V. Molodtsov, A.S. Proskuryakov, 
V.I.~Savrin and G.M.~Zinovjev are gratefully acknowledged.

\end{document}